# Analysis of Non-Linear Mode Coupling of Cosmological Density Fluctuations by the Pseudo-Spectral Method

Naoteru Gouda

*Department of Earth and Space Science,*
*Faculty of Science, Osaka University,*
*Toyonaka, Osaka 560, Japan*
*E-mail: gouda@vega.ess.sci.osaka-u.ac.jp*

## ABSTRACT

The pseudo-spectral method is proposed for following the evolution of density and velocity fluctuations at the weakly non-linear stage in the expanding universe with a good accuracy. In this method, the evolution of density and velocity fluctuations is integrated in the Fourier Space with using FFT. This method is very useful to investigate accurately the non-linear dynamics in the weakly non-linear regime. Because the pseudo-spectral method works directly in the Fourier space, it should be especially useful for examining behavior in the Fourier domain, for an example, the effects of the non-linear coupling of different wave modes on the evolution of the power spectrum. I show the results of this analysis both in one and three dimensional systems.

# I. INTRODUCTION

The formation of the large-scale structures is one of the most important problems which are unresolved still now in cosmology. It is commonly considered that small density fluctuations originated at the very early time grow to the observed structures due to the gravitational instability[1]. Then it is important to analyze the growth of density fluctuations in the expanding universe. Especially it is necessary to understand well the non-linear dynamics of density fluctuations after the decoupling time in order to clarify the mechanism of the large-scale structure formations. In analyzing the growth of density fluctuations, the statistical quantities like the power spectrum are often used[1]. As well known, the power spectrum tells us how large the amplitude of the density fluctuations on a scale is at any time. Furthermore, the power spectrum is converted to the two-point correlation function by the Fourier transformation[1]. So the growth rate and change of the shape of the power spectrum as time goes give us the important informations about the structure formations. The growth rate is well understood at the stage of the linear growth: after the decoupling time, the power spectrum grows in proportion to the square of the scale factor, $a^2$, independently of the wavenumber if we consider the matter dominated Einstein-de Sitter universe( the present density parameter $\Omega_0 = 1.0$), and then the shape of the power spectrum does not change as time goes in this stage. At the non-linear stage, however, the different parts of the power spectrum couple, that is, the non-linear coupling of the different wave modes occurs. This fact implies that the shape of the power spectrum changes as time goes and so the growth rate of the power spectrum depends on the wavenumber. This effect of mode coupling is the important basic process to determine the non-linear growth of density fluctuations. It is convenient to analyze the evolution equations of the density fluctuations in the Fourier space rather than in the real space in order to study the non-linear coupling of different modes. Of course, the analysis in the real space is also important and many works in the real space have been done by N- body simulations. I believe that the both analysis in the Fourier space and the real space are complementary. As for an example, the analysis in the real space is convenient for investigating the real pattern of collapsed objects. However, it is better to study in the Fourier space in order to clarify directly the effects of the non-linear mode coupling in the evolution of the power spectrum, as shown in details later.

Some works have been done for the non-linear growth of the power spectrum in the Fourier space[2,3]. They evaluated the second order contributions to the power spectrum. Bhunvnesh and Betschinger[3] analyzed the non-linear mode coupling in the second order perturbations systematically and showed some interesting features of transformation of the power between different modes. In this paper, we propose the pseudo-spectral method for analyzing the non-linear mode coupling. In this method, the evolution of density and velocity fields is integrated in the Fourier space with using the Fast Fourier Transformation(FFT) until the orbit-crossing appears on small scales. This method works directly in the Fourier space and so it should be very useful to evaluate the non-linear mode coupling. Furthermore, in this method, full order contributions to the power spectrum can be evaluated. The pseudo-spectral method works well in the weakly non-linear stage when the density fluctuations are not so much small and so higher order contributions can not be neglected in some cases though the orbit-crossing has not yet appeared on any scale. So in considering the full order contributions we can estimate the mode coupling in any case even with the initial power spectrum which has the sharp cut off, such as the power spectrum for hot dark matter models while the analyis in only using second order perturbation might not be correct enough in this case. However in the pseudo-spectral method the dynamic range of the power spectrum is, of course, restricted because of the constrain from the memory and CPU time of the computer rather than in the second order perturbation. Then I believe that both analysis by the pseudo-spectral method and the second order perturbation are complementary.

I analyze the effects of the mode coupling in both 1 and 3-dimensional systems by the pseudo-spectral method and also compare the numerical results with those given by the second order perturbation.

The pseudo-spectral method is briefly reviewed in Sec.II. The numerical results for 1 and 3-dimensional systems are shown in Sec.III. Finally, Sec.IV is devoted to conclusions and discussions.

## II.PSEUDO-SPECTRAL METHOD

The detailed explanation of the pseudo-spectral method(hereafter, we call it the PS method) is shown in Gouda[4]. Then I briefly review the PS method in this section.

The basic equations for the evolution of the density and velocity fields(continuity, Euler and Poisson equations) in the Fourier space are given as follows:

$$\frac{d\rho_{\mathbf{k}}(t)}{dt} + i\mathbf{k} \cdot \sum_{\mathbf{k}=\mathbf{k}'+\mathbf{k}''} \rho_{\mathbf{k}'}(t)\mathbf{v}_{\mathbf{k}''}(t) = 0$$

$$\frac{d\Phi_{\mathbf{k}}(t)}{dt} + \frac{1}{2}\sum_{\mathbf{k}=\mathbf{k}'+\mathbf{k}''} \mathbf{v}_{\mathbf{k}'}(t) \cdot \mathbf{v}_{\mathbf{k}''}(t) = -\Psi_{\mathbf{k}}$$

$$\Psi_{\mathbf{k}} = \begin{cases} -G\rho_{\mathbf{k}}/|\mathbf{k}|^2 & (\mathrm{k}=|\mathbf{k}|\neq 0) \\ 0 & (\mathrm{k}=0) \end{cases}$$

$$\mathbf{v}_{\mathbf{k}}(t) = i\mathbf{k}\Phi_{\mathbf{k}}(t) \qquad (1)$$

Here $\rho_{\mathbf{k}}, \mathbf{v}_{\mathbf{k}}, \Phi_{\mathbf{k}}$, and $\Psi_{\mathbf{k}}$ are the Fourier spectra for comoving density, peculiar velocity, velocity potential and comoving gravitation potential, respectively. The wavenumber $\mathbf{k} = (k_x, k_y, k_z)$ is truncated at $|k_x|, |k_y|, |k_z| < k_{max}$. The requirement that the fields be real in real space implies that the Fourier component $-\mathbf{k}$ is the complex conjugate of the Fourier component $\mathbf{k}$, e.g. $\rho_{-\mathbf{k}} = \rho_{\mathbf{k}}^*$. In deriving these equations, we assume that the velocity field has no vorticity because the rotational mode of the velocity is a decaying mode before the orbit-crossing appears. We can therefore write $\mathbf{v}(\mathbf{x}) = \nabla \Phi(\mathbf{x})$, where $\mathbf{v}$ is the velocity field and $\Phi$ is the velocity potential. In this paper, we consider only the case that the present density parameter $\Omega_0 = 1$ and the cosmological term $\Lambda = 0$. Then the relation between time $t$ in Eq.(1) and redshift $z$ is simply given by $t = -\frac{2}{H_0}\sqrt{1+z}$, where $H_0$ is the present hubble constant[4]. Equations in (1) are differential equations, so we can integrate them accurately with, say a 4-th order Runge-Kutta scheme.

In calculating the non-linear coupling terms in Eq.(1), $\sum_{\mathbf{k}=\mathbf{k'}+\mathbf{k''}} \rho_{\mathbf{k'}}(t)\mathbf{v}(\mathbf{k''},t)$ and $\sum_{\mathbf{k}=\mathbf{k'}+\mathbf{k''}} \mathbf{v}_{\mathbf{k'}}(t) \cdot \mathbf{v}_{\mathbf{k''}}(t)$, we use FFT. First, we transform $\rho_{\mathbf{k}}(t)$ and $\mathbf{v}_{\mathbf{k}}(t)$ to the real-space fields, $\rho(\mathbf{x},t)$ and $\mathbf{v}(\mathbf{x},t)$, via FFT. Then we take the products $\rho(\mathbf{x},t) \cdot \mathbf{v}(\mathbf{x},t)$ and $\mathbf{v}(\mathbf{x},t) \cdot \mathbf{v}(\mathbf{x},t)$. Finally, we transform these products back into Fourier space to obtain the coupling terms in Eq.(1). This method is called the "pseudo-spectral method (PS method)". This method is often used in the fields of fluid dynamics[5]. The computation time for numerical calculations is proportional to $\sim N \ln N$, where $N$ is $(2k_{max})^3$, because of the FFT, though it would be proportional to $\sim N^2$ if we calculated the non-linear coupling terms directly.

Here, however, we must note that there exists a serious problem in calculating numerically the non-linear coupling terms by using FFT, according to the procedure mentioned above: the problem of aliasing. There appear unphysical Fourier spectra in this procedure because we must cut the Fourier spectrum at the finite wavenumber, $k_{max}$. However we can remove this alias. The detailed explanation of the aliasing problem and also the procedure to remove the alias are shown in Gouda[4]. In using this procedure, we do not worry about the aliasing problem.

As a numerical test of the PS method[4], I compared the results obtained by the PS method with those obtained by other methods. For 3-dimensional systems, I calculated the time-evolution of the power-spectrum by the particle-mesh N-body simulation(CIC method) and compared the results with those obtained by the pseudo-spectral method with the same initial conditions as in CIC. I found that both results are almost in agreement with each other in the weakly non-linear stage when the density fluctuations are not so much small and so higher order contributions can not be neglected in some cases though the orbit-crossing has not yet appeared on any scale.

I also found that in one-dimensional systems the results obtained by the pseudo-spectral method are in good agreement with the exact solution [6] for the time-evolution of the density fields, which exist only in one-dimensional systems, prior to the first appearance of orbit-crossing.

Furthermore I have tested the time reversibility in following the time evolution of density fields and found that the accuracy for the time reversibility is very good in the PS method.

## III. NUMERICAL RESULTS

As a first step, I analyze the effects of the mode coupling in the 1- dimensional system. In this system, we know the exact solution for the evolution of the density fields until the first orbit-crossing occurs, that is called, Zel'dovich solution[6]. By comparing the Zel'dovich solution with the numerical results by the PS method, I found that the PS method traces the evolutions of the density fields very accurately.

In order to study the effect of the mode coupling, I divide the non- linear coupling term into three parts:

$$\sum_{k=k'+k''} \rho_{k'}(t) v_{k''}(t) =$$

$$\sum_{k>k'>0} \rho_{k'} v_{k-k'} \quad (A)$$

$$+ \sum_{k'>0} \rho_{-k'} v_{k+k'} + \sum_{k'>0} \rho_{k+k'} v_{-k'} \quad (B)$$

$$+ \rho_0 v_k \quad (C)$$

The term (A) is the contribution of the power on smaller wavenumber $k - k' < k$, to the power at $k$, that is, long-wave mode. And the terms in (B) are the contribution of the power from higher wavenumber $k + k'$ to the power at $k$, short-wave mode. The final term (C) is the linear contribution. We can study the effects of mode coupling in each term, separately. This estimate can be done only by the analysis in the Fourier space, not in the real space by N-body simulations or using Zel'dovich solution.

I performed the following numerical calculations. As for initial conditions, we assume that at the initial redshift $z_{in}$ the power spectrum $P(k, z_{in}) = |\rho_k|^2$ is in pro-portion to $k^n$ and $n$ is integer ranging from -2 to 5(case I). And also I consider only the case in which the initial power spectrum $P(k, z_{in}) \propto k (k \leq k_{cut} = 10), P(k) = 0 (k > k_{cut} = 10)$(case II). I consider the case in which $k_{max} = 64$ and $z_{in} = 99$. The normalization of the amplitude of the initial power spectrum is determined by the criterion that the first orbit-crossing would artificially appear at $z = -0.2$. Therefore the orbit-crossing does not appear on any scale under consideration until $z = 0$, so that we can apply the PS method until $z = 0$ without worrying about the orbit-crossing. The time evolution of the power spectra are shown in Figs.1(a) ∼ 1(e) for $n = -2, -1, 0, 2$ and 5, respectively. The power spectra at $z = 99, 30, 10, 2$ and 0 are shown in each figure. The solid curves represent the "exact" power spectra which are calculated including the full contributions, (A)+(B)+(C), in the non-linear coupling term. The dotted curve represents the power spectrum at $z = 0$, which is calculated as follows: until $z = 2$, the power spectrum is calculated with full contributions in the non-linear coupling term. After $z = 2$, however, the contribution from smaller $k$ (term(A)) and the linear contribution(term(C)) are only included in following the evolution of the power spectrum, that is, we remove the contribution from higher $k$(term(B)) (we call it "long-wave"). The dotted and short-dashed curve represents the power spectrum at $z = 0$, which is calculated including the only contribution from higher $k$(term(B)) and the linear contribution (term(C)) after $z = 2$(we call it "short-wave"). In this case I do not show the power spectrum at $z = 0$ on higher $k$ in figures because there appears the artificial errors near $k_{max}$ due to the finite resolution in the wavenumber space. The dotted and long-dashed curve represents the power spectrum at $z = 0$ given by the calculation including only the linear contribution (term(C)) after $z = 2$. Then in this case the shape of the power spectrum at $z = 0$ is the same as one at $z = 2$.

In comparison of the "exact" power spectrum at $z = 0$ with the other ones, such as "long-wave", "short-wave" and the linear one, we find that the contribution from smaller $k$ [term(A): long-wave mode] is positive one for $n \lesssim -1$ because the "long wave" (including terms (A)+(C)) exceeds the linear power spectrum(including only the term(C)). On the other hand, the contribution from higher $k$[term (B): short-wave mode] is negative because the "short-wave"(including terms (B)+(C)) is smaller that the linear one. Furthermore we can see that for the case in which $n \lesssim -1$, the "exact" power spectrum at $z = 0$ is larger than the linear one on higher $k$ because the positive contribution of the long-wave mode exceeds the negative contribution of the short-wave mode. This fact implies that the growth rate is larger than the linear growth rate on higher $k$. These results are consistent with the results by the second order perturbation[3]. As for $0 \lesssim n \lesssim 4$, the "exact" power spectrum , "long-wave" and "short-wave" at $z = 0$ agrees well with the linear one. This implies that the non-linear couplings( long-wave mode and short-wave mode) do not contribute so much to the evolution of the power spectrum. The reason is that the amplitude of the linear contribution on a scale is larger than that of the non-linear coupling terms contributed from the other scales because of the shape of the power spectrum and the over all relatively small amplitude of the power spectrum at $z \geq 0$ due to the condition of the normalizaion.

As for $n \gtrsim 5$, however, the contribution from higher $k$, short-wave mode, turns to be positive and dominate on higher $k$. The reason is as follows: in this case, the amplitude of the power spectrum on the smaller $k$ scale is much less than that on the higher $k$ scale . And so the non-linear coupling term contributed form the higher $k$ scales, short-wave mode, contribute more dominantly than the linear term on the smaller $k$ scale althogh the amplitude of the power spectrum on all scales *itself* is much less than unity. Moreover as for the case (II) of the initial power spectrum, we can see that the same features as for $n \lesssim -1$ appear on the scales above $k_{cut}$. This case can be evaluated in considering the full order contribution by the PS method while the analysis in using only the second order perturbation might not be correct enough in this case.

Next I consider the 3-dimensional system. In this system, we consider the case in which $k_{max} = 15$. I found that the qualitative features of the following results do not change so much when the different $k_{max}$ is considered. Figures 2(a) $\sim$ 2(f) show the time evolutions of the power spectra for $n = -4, -2, -1, 2, 5$ and 7, respectively. The meanings of some kinds of curves are the same as in Figs.1 though I do not show the linear contribution (dotted and long-dashed curve in Figs.(1)) in Figs.2. From these figures, we find that the same features as in the 1-dimensional system appear: as for $n \lesssim -3$, the contributions from higher $k$, short-wave mode, is negative. On the other hand, the contribution from smaller $k$, long-wave mode, is positive and dominates on higher $k$. As for $-2 \lesssim n \lesssim 4$, the non-linear contributions are smaller than the linear contribution. The reason is the same as in the one-dimensional system. These results are also consistent with the results obtained by the second order perturbation[3]. As for $n \gtrsim 5$, the contribution from higher $k$ turns to be positive and dominates on smaller $k$. The reason is the same as in the one-dimensional system. This result might be related to the minimal fluctuations[1] and is suggested by the previous work[7].

## IV.CONCLUSIONS AND DISCUSSIONS

The PS method is very useful to investigate accurately the non-linear dynamics in the weakly non-linear regime when the density fluctuations are not so much small and so higher order contributions can not be neglected in some cases though the orbit-crossing has not yet appeared on any scale. Especially, it is convenient to examine the effects of the non-linear coupling of the different wave modes on the evolutions of the power spectrum because the PS method works directly in the Fourier space and can include the full order contributions in the non-linear coupling terms. We find in both the 1 and 3-dimensional systems that the contribution from smaller $k$ in the non-linear coupling term, long-wave mode, is positive and dominates on higher $k$ when the initial power index $n$ is smaller. On the other hand, the contribution from higher $k$, short-wave mode, is negative. These results agree with the previous works by the second order perturbation for the power-law power spectrum without the sharp cut off [2,3]. Furthermore we could show that these results are qualitatively applicable to the case that the initial power spectrum has the sharp cut off although the second order perturbation might not be able to allpy to this case. As $n$ increases, the contribution from the non-linear coupling decreases and the linear contribution becomes to dominate on any scale. And the contribution from higher $k$ turns to become positive and dominate on smaller $k$ when $n$ is much larger. It is interesting that the above features are similar in both the 1 and 3-dimensional systems. Moreover it must be noted that the initial power spectrum with $n_3 \lesssim 0$ in 3-dimensional system has the same features for the effects of contributions in the non-linear coupling term as the initial power spectrum with $n_1 \sim n_3 + 2$ in the 1-dimensional system. That is, the power spectrum with $n_3$ in the 3-dimensional systems has the similar evolution of the power spectrum with $n_1$ in the 1-dimensional systems. For an example, the contributions of the non-linear modes become negligible $n_3 \gtrsim -2$ and $n_1 = n_3 + 2 \gtrsim 0$. This fact might be resulted from that the difference of the dimension of the volume element in the coupling term $\sum_{k=k'+k''} \rho_{k'}(t) v_{k''}(t)$ in the 1 and 3-dimensional system is two. As for $n \gtrsim 5$, however, the contribution from higher $k$, short-wave mode, in both 1 and 3-dimensional systems is positive and dominates on smaller $k$. This fact might be related to the minimal fluctuations with $n = 4$[1, 7].

After when we face on the strongly non-linear regime where the orbit-crossings occur on smaller scales, we can not use the evolution equations (1), that is, the PS method can not be applied to these scales and we must perform N-body simulations. It is very interesting to study further the non-linear dynamics at the strongly non-linear regime in details while some interesting results have been reported[8].

## ACKNOWLEDGMENT

I am very grateful to T.Tsuchiya and referees for useful comments. This work was supported in part by the Grant-Aid for Scientific Research Fund from the Ministry of Education, Science and Culture of Japan,
Nos.05302015 and 06640352.

# FIGURE CAPTIONS

FIG.1(a). Time evolution of the power spectrum with the initial power index $n = -2$ in the 1-dimensional system. The power spectra at $z = 99, 30, 10, 2$ and 0 are shown by the solid curves. The dotted curve represents the power spectrum at $z = 0$, "long-wave", which is calculated after $z = 2$ in considering only the contribution of the non-linear coupling from long-wave mode and the linear contribution. The dotted and short-dashed curve represents the power spectrum at $z = 0$, "short-wave", in considering only the contribution from the short wave-mode and linear contribution after $z = 2$. The dotted and long-dashed curve represents the power spectrum at $z = 0$ in considering only the linear contribution after $z = 2$.

FIG.1(b). The same as FIG.1(a), but for the initial power spectrum with $n = -1$.

FIG.1(c). The same as FIG.1(a), but for the initial power spectrum with $n = 0$.

FIG.1(d). The same as FIG.1(a), but for the initial power spectrum with $n = 2$.

FIG.1(e). The same as FIG.1(a), but for the initial power spectrum with $n = 5$.

FIG.1(f). The same as FIG.1(a), but for the case(II) as the initial power spectrum.

FIG.2(a). Time evolution of the power spectrum with the initial power index $n = -4$ in the 3-dimensional system. The solid, dotted, dotted and short-dashed curves have the same meanings as in Figs.1

FIG.2(b). The same as FIG.2(a), but for $n = -2$.

FIG.2(c). The same as FIG.2(a), but for $n = -1$.

FIG.2(d). The same as FIG.2(a), but for $n = 2$.

FIG.2(e). The same as FIG.2(a), but for $n = 5$.

FIG.2(f). The same as FIG.2(a), but for $n = 7$.

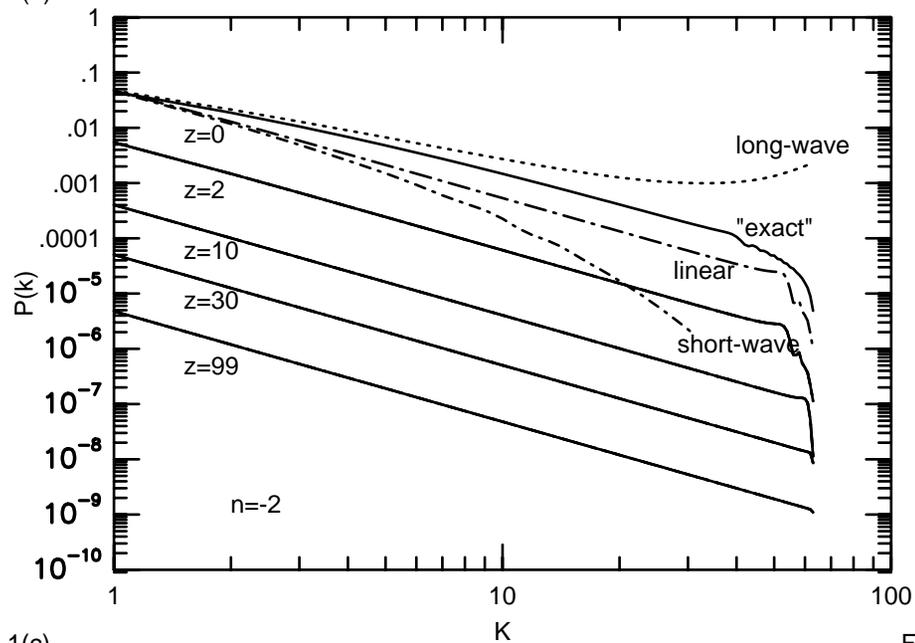 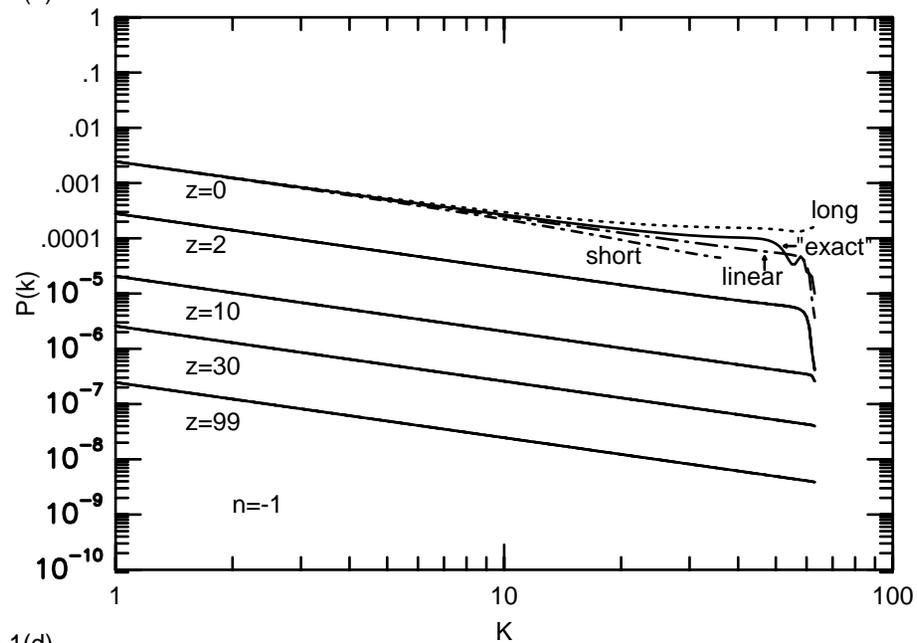 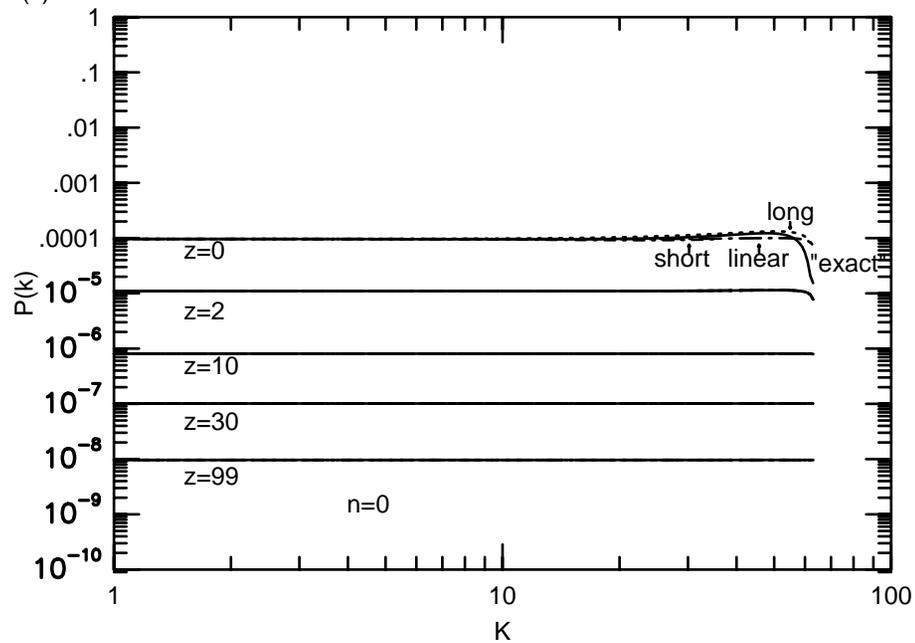 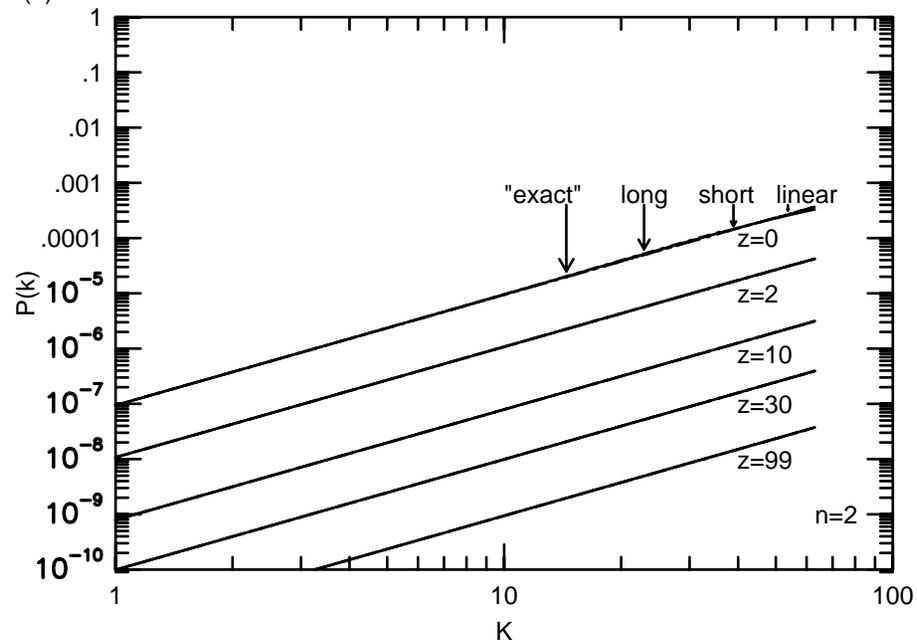

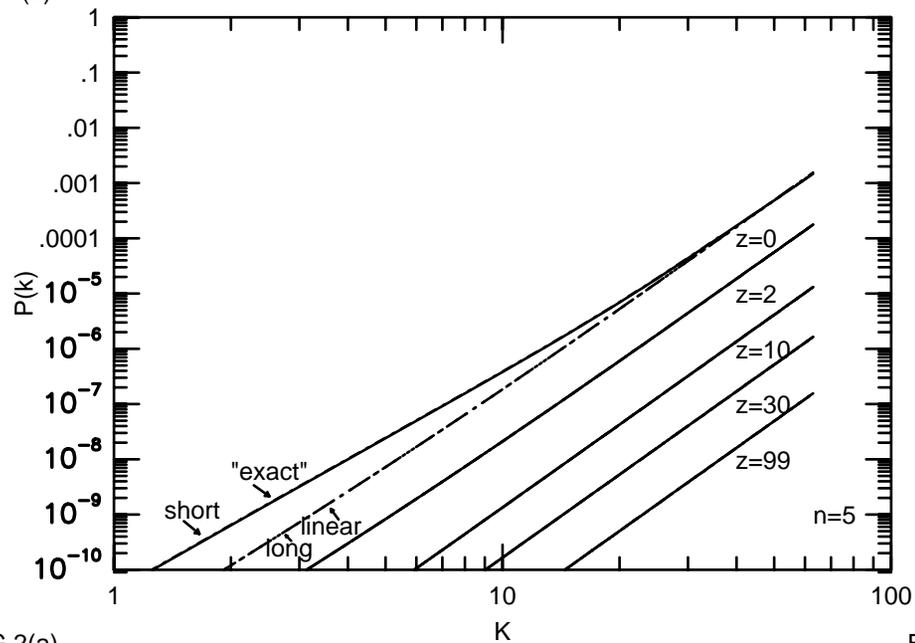
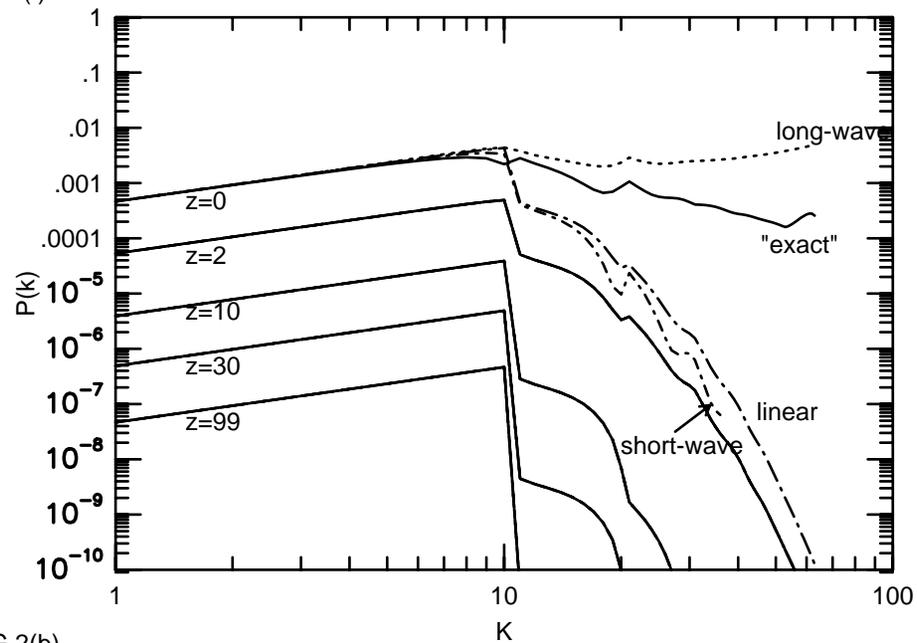
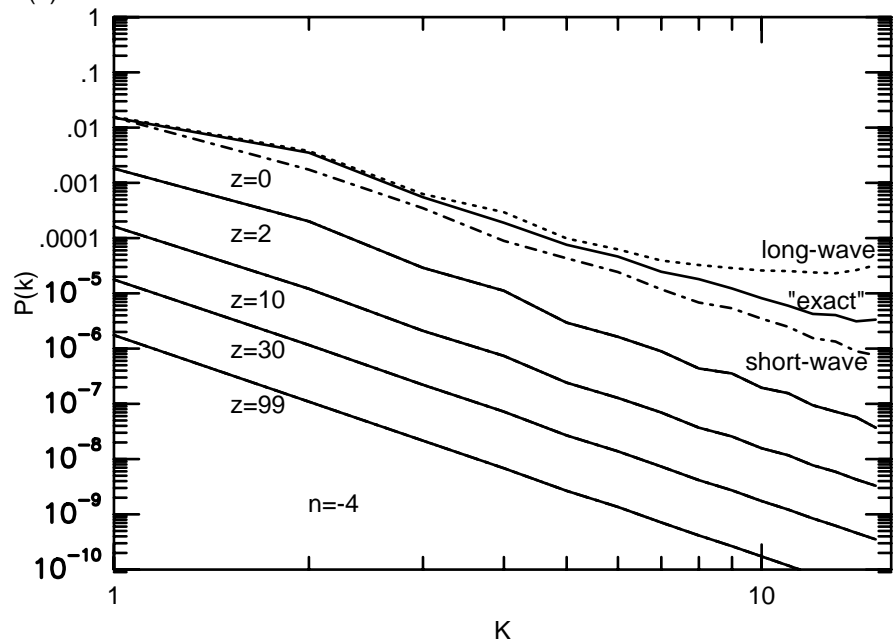
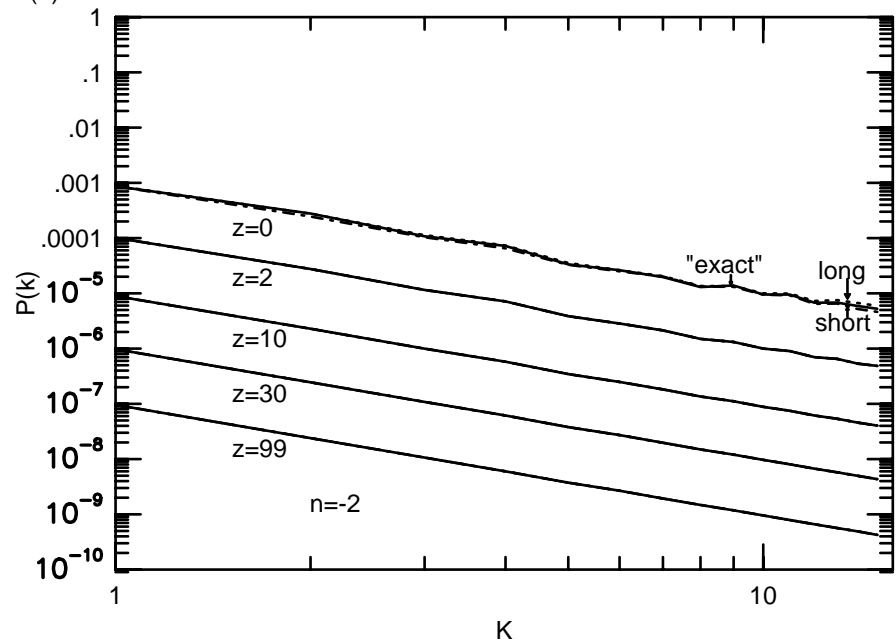

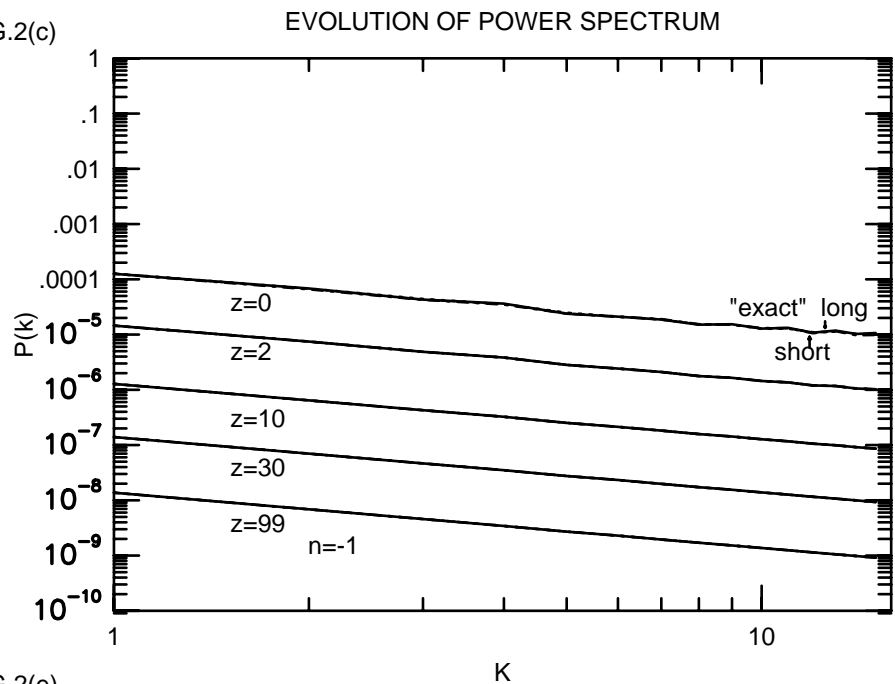
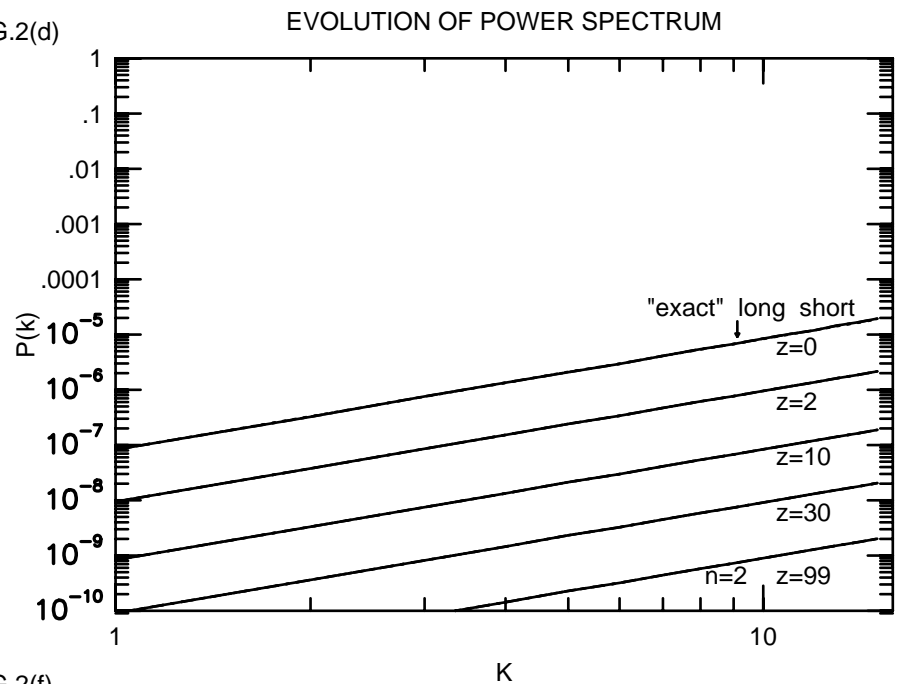
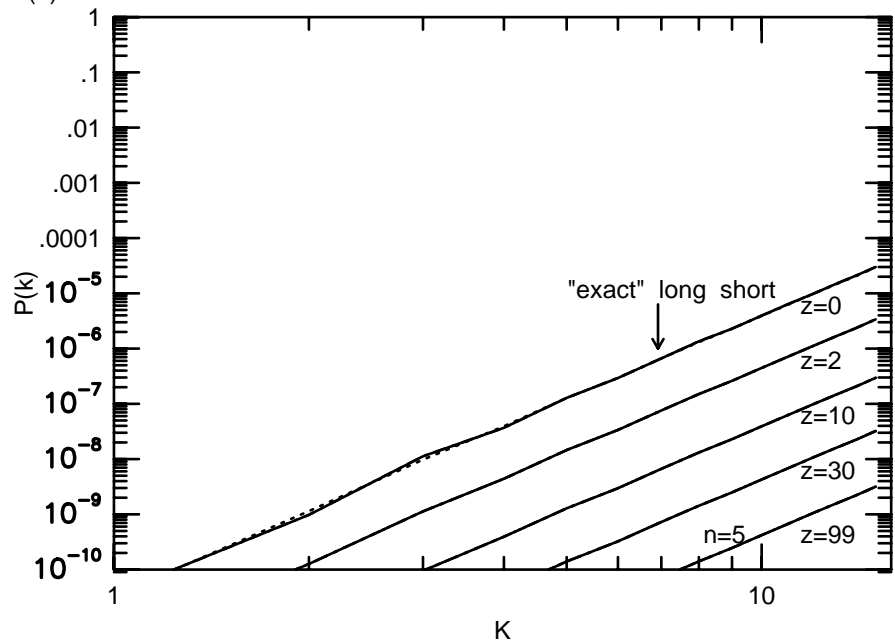
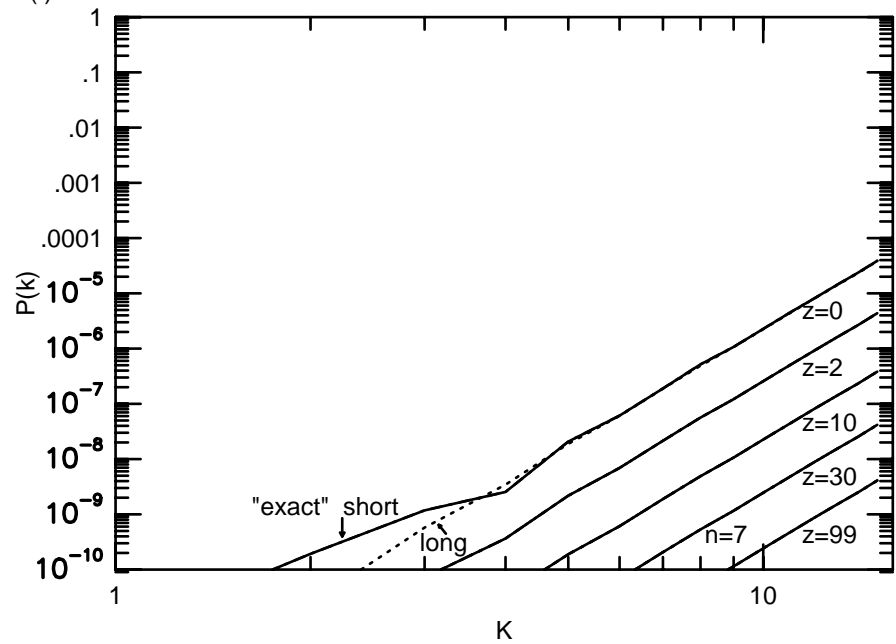